\documentclass[10pt,twocolumn,letterpaper]{article}

\usepackage{iccv}              

\definecolor{iccvblue}{rgb}{0.21,0.49,0.74}
\usepackage[pagebackref,breaklinks,colorlinks,allcolors=iccvblue]{hyperref}

\newcommand{\new}[1]{\textcolor{black}{#1}}
\newcommand{\fin}[1]{\textcolor{black}{#1}}

\usepackage{hyperref}
\usepackage[ruled]{algorithm2e} 

\SetAlFnt{\small}
\SetAlCapFnt{\small}
\SetAlCapNameFnt{\small}
\SetAlCapHSkip{0pt}

\usepackage{amsmath}
\usepackage{textcomp} 
\usepackage{enumitem}  
\usepackage{nicefrac}
\usepackage{siunitx}

\def\paperID{8584} 
\def\confName{ICCV}
\def\confYear{2025}

\title{Augmented Mass-Spring Model for Real-Time Dense Hair Simulation}

\author{
J. Alejandro Amador H.\\
KAUST\\
{\tt jorge.amadorherrera@kaust.edu.sa}
\and
Yi Zhou \quad Xin Sun \quad Zhixin Shu\\
Adobe Research\\
{\tt \{yizho, xinsun, zshu\}@adobe.com}
\and
Chengan He\\
Yale University\\
{\tt chengan.he@yale.edu}
\and
S\"oren Pirk\\
Kiel University\\
{\tt sp@informatik.uni-kiel.de}
\and
Dominik L. Michels\\
KAUST\\
{\tt dominik.michels@kaust.edu.sa}
}

\begin{document}
\twocolumn[{
\maketitle
\begin{center}
    \captionsetup{type=figure}
    \includegraphics[width=\textwidth, keepaspectratio]{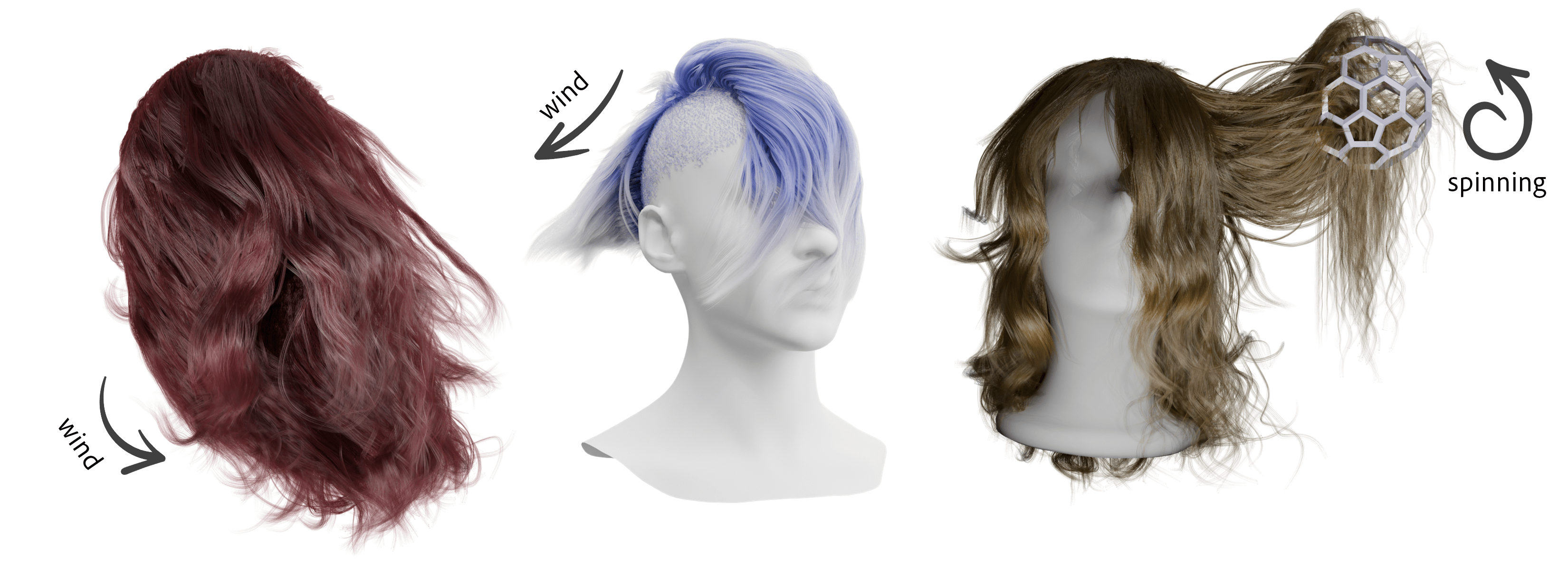}
    \caption{Our AMS model for dense hair simulation is capable of handling complex settings with a large number of strands, efficiently simulating scenarios such as hair under wind (left, $14,718$ strands at $67$ FPS), hair-face contact (middle, $7,528$ strands at $156$ FPS), and hair interacting with complex objects (right, $10,298$ strands at $114$ FPS), all in real time.}
    \label{fig:teaser}
\end{center}
}]
\begin{abstract}
We propose a novel Augmented Mass-Spring (AMS) model for real-time simulation of dense hair at the strand level. Our approach considers the traditional edge, bending, and torsional degrees of freedom in mass-spring systems, but incorporates an additional one-way biphasic coupling with a ghost rest-shape configuration. Through multiple evaluation experiments with varied dynamical settings, we show that AMS improves the stability of the simulation in comparison to mass-spring discretizations, preserves global features, and enables the simulation of non-Hookean effects. Using a heptadiagonal decomposition of the resulting matrix, our approach provides the efficiency advantages of mass-spring systems over more complex constitutive hair models, while enabling a more robust simulation of multiple strand configurations. Finally, our results demonstrate that our framework enables the generation, complex interactivity, and editing of simulation-ready dense hair assets in real time. 
\end{abstract}    
\section{Introduction}
\label{sec:introduction}
Hair dynamics play a crucial role in enhancing the visual realism of digital characters. However, simulating hair remains one of the most challenging and resource-intensive tasks, due to its thin, flexible structure, the interactions between strands (including friction and collisions with both the body and other hair strands), and the sheer volume of hair, which can range from $100,000$ to $200,000$ strands on a single head. Over the years, various techniques have been developed to replicate hair dynamics, including sophisticated models like the Discrete Elastic Rods (DER) approach~\cite{bergou2008discrete}. However, due to the high computational demands, simulations using physics-based models are typically limited to hundreds of strands for low-latency applications, which restricts their uses in games and animation. To achieve interactive performance, many systems simulate a smaller number of guide strands and apply linear or neural interpolation techniques 
to upsample the strand count in real time. While this approach improves performance, it often sacrifices fidelity, particularly in areas with complex strand collisions.

In recent efforts to simulate a larger number of strands in real time, more efficient integration routines~\cite{daviet2023interactive} that parallelize DER-based physics models have emerged, alongside data-driven methods using neural networks to predict hair dynamics. However, these methods face robustness issues in handling outlier shapes and dynamic settings and dealing with complex interactive scenes and intricate hairstyles. 
In this paper, we propose a novel Augmented Mass-Spring (AMS) model for real-time dense hair simulation. Standard Mass-Spring (MS) model~\cite{jiang2020real} is methodologically simpler and computationally cheaper than Discrete Elastic Rod (DER) models or neural-based methods. However, by design, the MS model struggles to capture global strand behavior, leading to issues such as instability, excessive sagging, and a loss of hair structure during simulation. To address these limitations, we introduce several key augmentations to the MS model. For improving the stability of the system and encoding global strand features, we design a novel but simple scheme. We incorporate one-way biphasic interactions, combined with a ghost rest-shape, applied to the particles along the hair strands (Figure~\ref{fig:diagram_spring_systems}). This approach stabilizes the simulation while preserving global structural integrity. To maintain the dynamic realism and computational efficiency of MS models, we design a two-stage hybrid Eulerian/Lagrangian scheme that integrates fluid system and particle system to handle hair interactions. This combination allows us to simulate intricate hair behavior without sacrificing performance.

\begin{figure}
\centering
\includegraphics[width=\columnwidth, keepaspectratio]{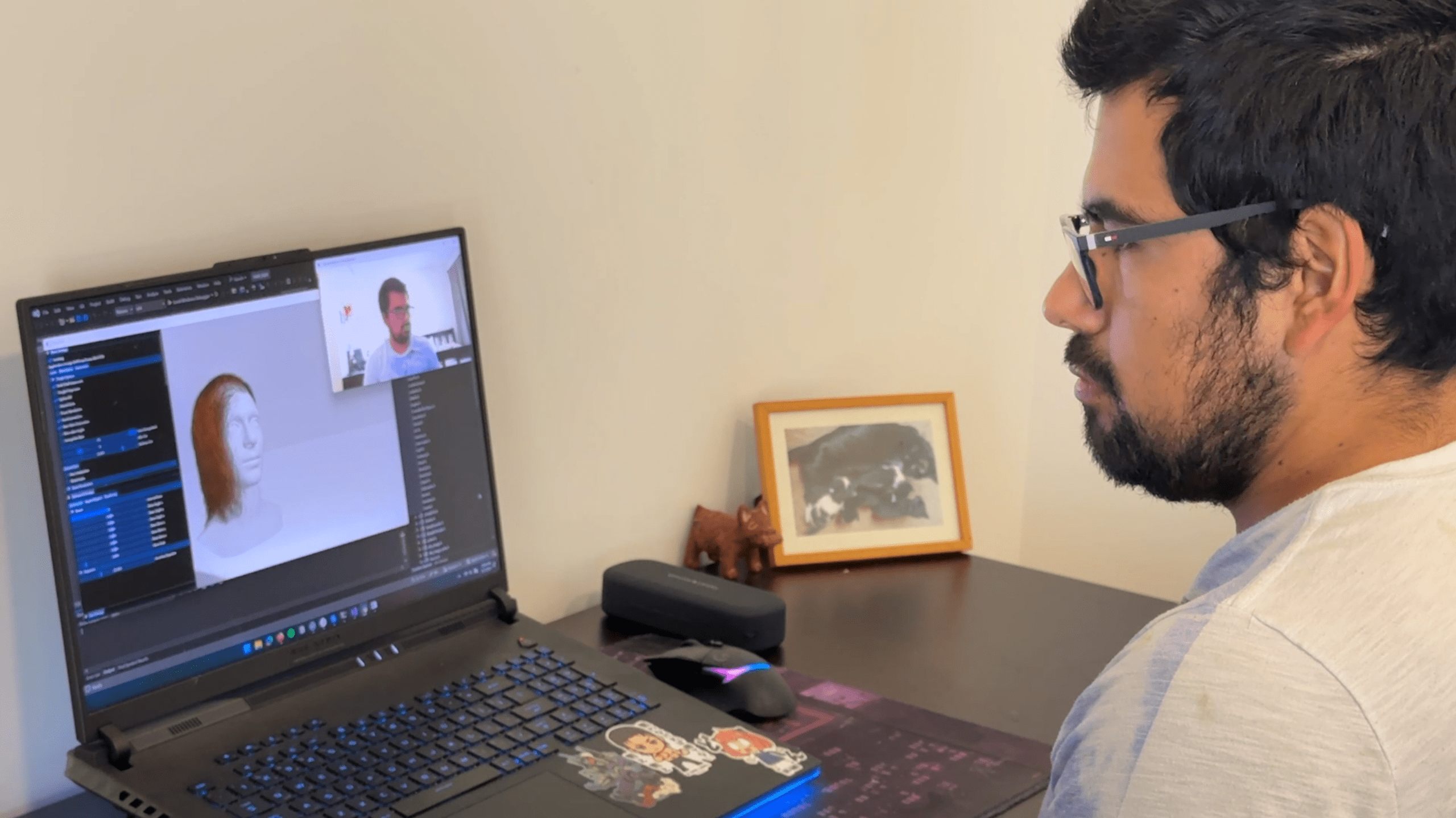}
\caption{Integrating video-based facial tracking \cite{deng2019retinaface} with our real-time hair simulation enables the control of digital avatars with dynamic and coherent hair motion.}
\label{fig:facial_tracking}
\end{figure}

We evaluate our framework through comparative experiments across various dynamic and interaction settings, and additionally showcase \textbf{strand-level interactive simulations and grooming of dense hair} on a consumer-grade computer, utilizing less than 1 gigabytes of GPU memory. To the best of our knowledge, AMS is the first framework to enable real-time simulation of a wide range of hair and facial hair styles, including curly hair, asymmetrical styles, and ponytails, while having the capability of capturing \textbf{fine collision details} when interacting with complex geometries (Figure~\ref{fig:teaser}) and \textbf{handling extreme forces}, such as those encountered during intense motion, like on a roller coaster (Figure~\ref{fig:roller_coaster}). Furthermore, by integrating video-based facial tracking with our real-time hair simulation, we enable the control of digital avatars with dynamic and coherent hair motion (Figure~\ref{fig:facial_tracking}).
\section{Related Work}
\label{sec:related_work}
Here we discuss the physics-based hair simulation model and the recently emerged neural-based approaches.
\subsection{Physics-based Approaches}
Physics-based hair simulation methods vary in the geometric and constitutive models they use to simulate either individual hair strands or clumps of hair. Approximating hair as large bundles has been explored through various techniques, including cubic lattice representations \cite{volino2006real}, volumetric models \cite{lee2018skinned, wu2016real}, and 2D strips \cite{koh2001simple}. While these approaches are typically very efficient, they are limited in their ability to capture effects and interactions that require modeling at the level of individual strands. In contrast, individual hair strands have been modeled using mass-spring systems \cite{rosenblum1991simulating} and multi-body chains \cite{chang2002practical}. Some approaches, such as the use of lattice deformers with additional springs \cite{selle2008mass}, enhance system stability, while others focus on efficient solvers and improved hair-body interactions \cite{jiang2020real}.

More physically accurate and complex models, such as the super-helix scheme \cite{bertails2006super}, DER \cite{bergou2008discrete}, and exponential time integrators \cite{michels2015physically, 10.1145/3072959.3073706}, provide greater realism but at the cost of significantly higher computational demands. Recently, Daviet \etal \cite{daviet2023interactive} made significant engineering advancements to accelerate the DER model by parallelizing strand computations, enabling the simulation of over $10,000$ guide strands in real time. This approach significantly improves the fidelity of real-time hair simulation. However, despite these optimizations, DER-based methods remain computationally complex for simulating strand dynamics, making them challenging to apply in more intricate interactive scenarios with complex hairstyles. Additionally, these methods require specialized integration routines to address sagging artifacts \cite{hsu2023sag}.

In this context, we propose AMS as an evolution of traditional mass-spring schemes, enabling more robust and diverse hair dynamics simulations. This approach improves the fidelity of mass-spring-based hair simulations while preserving the efficiency advantages of these models.

\subsection{Neural-based Approaches}
With the rise of machine learning in physics, recent research has begun exploring neural approaches to predict hair dynamics using data generated by physics simulators, with the aim of accelerating animation. One line of work focuses on predicting individual hair strand shapes based on gravity direction and head pose~\cite{zhou2023groomgen}. Another approach uses deep learning to add fine details to interpolated dense hair from sparse guide strands~\cite{lyu2020real, shen2023ct2hair}. \fin{More recently, Quaffure~\cite{stuyck2025quaffure} introduced a real-time neural framework for quasi-static hair simulation that achieves plausible dynamic responses via a self-supervised loss. Additional efforts explore vision-based reconstruction hair, such as HairStep~\cite{zheng2023hairstep}, which introduces an intermediate representation consisting of a strand map and depth map, and the two-stage method of Neural Haircut~\cite{sklyarova2023neural}, where coarse geometry is first recovered using volumetric representations, followed by strand-level refinement. Further advances, like NeuWigs~\cite{wang2023neuwigs}, leverages a 3D volumetric autoencoder for hair capture and animation, while MonoHair~\cite{wu2024monohair} separates hair modeling into exterior and interior geometries to achieve high-fidelity hair reconstruction from a monocular video.}

While these methods can achieve real-time results, they are heavily dependent on the distribution of the training data, limiting their applicability to the data’s domain. Consequently, these models may struggle in scenarios outside their training data, leading to gaps in performance when faced with unfamiliar cases. In practice, it is infeasible to generate training data that covers all possible scenarios, given the vast combinations of internal parameters, external forces, and collision situations. Furthermore, the grooms produced by these methods lack embedded dynamic information, causing their behavior to potentially deviate significantly from the original intent during motion. In contrast, digital grooming within our dense hair simulation framework adheres to physical constraints even for manipulated strands, ensuring that the final groom will exhibit predictable dynamic behavior and handling complex hair-solid interactions.
\section{Formulation}
\label{sec:method}
In this section, we introduce our augmented mass-spring formulation, and describe how we handle hair-hair and hair-solid interactions.

\subsection{Augmented Mass-Spring Model}
\new{In a general MS scheme, each hair strand is discretized into particles connected by springs, providing edge, bending, and torsion degrees of freedom. However, issues such as ill-defined torsion springs and collapsed tetrahedra can arise when consecutive particles become degenerate. To address these challenges, Selle \etal~\cite{selle2008mass} introduced strongly coupled ghost particles and \emph{altitude} springs (Figure~\ref{fig:diagram_spring_systems}, middle), improving stability. However, this model relies on highly stiff springs to maintain local features and minimize jittering, which in turn introduces numerical divergence and limits the simulation time step size \cite{ward2007survey}. Furthermore, the purely local connectivity of the spring system leads to significant sagging at initialization and a loss of global hair structure.}

\new{To overcome these limitations, we introduce two weakly coupled interactions in our MS formulation (Figure~\ref{fig:diagram_spring_systems}, right). First, an \emph{angular} interaction to ensure nondegenerate tetrahedra by applying forces that prevent excessive collapse of consecutive particles. Second, an \emph{integrity} interaction that couples each particle to its corresponding ghost based on the total strand displacement, counteracting sagging and preserving the overall hair shape. This mechanism maintains global features independently of discretization density by offsetting the accumulated weight of particles. Moreover, to ensure these additional forces do not over-influence the system and decrease the simulation fidelity, we set their coupling constants several orders of magnitude lower than those of traditional local springs. A more detailed motivation is provided in our supplementary material.}
\subsubsection{Model Description}
\begin{figure}
    \centering
    \includegraphics[width=\columnwidth, keepaspectratio]{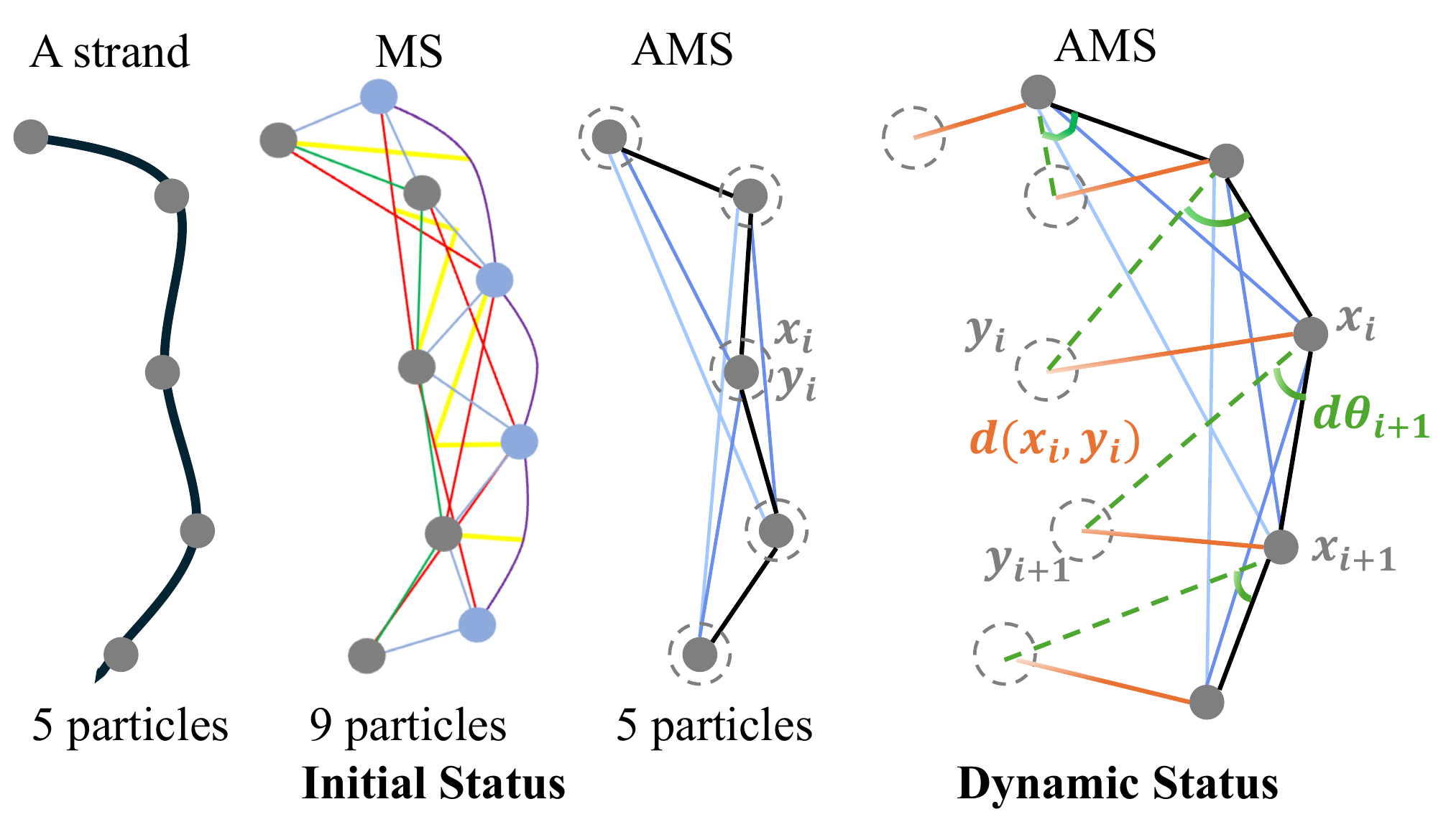}
    \caption{Schematic of a hair strand: initially discretized with particles (left); built as MS (middle) with edge (green), bending (blue), torsion (red), ghost (blue/purple), and altitude (yellow) springs; and as our AMS (right), using a ghost rest-shape (dotted circles) and one-way biphasic springs (purple outline). MS requires $2N{-}1$ coupled particles; AMS only uses $N$. Right: deviations from rest-shape are handled via integrity springs ($\kappa_I$) based on point distance, and angular springs ($\kappa_\alpha$) based on angle distance $d\theta$, forming biphasic links to the ghost.}
    \label{fig:diagram_spring_systems}
\end{figure}

We consider the basic edge, bending, and torsion springs, for which an uniform spring constant $\kappa_{L}$ is used, as well as  a weakly interaction with a ghost rest-shape via one-way springs that connect each particle with their corresponding ghost in the rest configuration, as demonstrated in Figure~\ref{fig:diagram_spring_systems}, right. Moreover, in general, MS needs $2N-1$ coupled particles, while AMS uses only the original $N$ particles. This difference between models makes AMS more efficient in terms of memory, as only a matrix with half the size of that in MS is needed to solve during each iteration.
\subsubsection{Biphasic Interaction}
Given a discretized strand containing $n$ particles with positions $\boldsymbol{x}_{0},...,\boldsymbol{x}_{n-1}$, we define $\boldsymbol{y}_{i}$ to be the position of the ghost corresponding to the $i-th$ particle. The first part of the interaction uses an \emph{integrity} spring with zero rest length and tension $T_{I}$ computed as
\begin{equation}
\label{eq:integrity_spring}
    T_{I} = \kappa_{I} d(\boldsymbol{x}_{i}, \boldsymbol{y}_{i})\,,
\end{equation}
where $\kappa_{I}$ is the spring constant, and $d\colon\mathbb{R}^{3}\times\mathbb{R}^{3}\to\mathbb{R}^{3}$ is the distance function between two vectors. This spring connects each particle to its corresponding ghost configuration. Moreover, note that, at initialization, $\boldsymbol{x}_{i} = \boldsymbol{y}_{i}, \, \forall i \in \{1,...,n\}$.  We also introduce an \emph{angular} spring, so that, if $d\theta_{i+1}$ is the angle between segments $\boldsymbol{x}_{i} \boldsymbol{y}_{i+1}$ and $\boldsymbol{x}_{i} \boldsymbol{x}_{i+1}$, then the tension of the spring $T_{\alpha}$ is given by
\begin{equation}
    T_{\alpha} = \kappa_{\alpha} d\theta_{i+1}\,,
\end{equation}
where the spring connects the $(i+1)th$ particle to its ghost. At initialization, all the edges are aligned to the ghost counterpart, so that $d\theta_{i+1} = 0 \, \forall i \in \{1,...,n\}$. Each pair of springs between ghost and real particles act in a biphasic fashion, \ie, they are connected in parallel.


\subsubsection{Linearization}
\label{sec:linearization}
We proceed as before, letting $\boldsymbol{x}^{n}_{i}$, $\boldsymbol{y}^{n}_{i}$, $\boldsymbol{v}^{n}_{i}$, and $\boldsymbol{w}^{n}_{i}$ denote, respectively, the position of each ghost-real pair, and their correspondent velocities. The superscripts denote the current time step. Furthermore, the vectors $\boldsymbol{X}^{n}, \boldsymbol{Y}^{n}, \boldsymbol{V}^{n}, \boldsymbol{W}^{n}\in\mathbb{R}^{3N}$ hold the position and velocities of the complete ghost-real strand pair. Using this notation, the backward Euler equations from time $t^{n}$ to $t^{n+1}$ can be written as
\begin{align}
    \boldsymbol{X}^{n+1} &= \boldsymbol{X}^{n} + \Delta t \boldsymbol{V}^{n+1}\,, \\
    \boldsymbol{V}^{n+1} &= \boldsymbol{V}^{n} + \Delta t \boldsymbol{M}^{-1} \left( \boldsymbol{F}^{n} + \boldsymbol{S}^{n+1} - \boldsymbol{G}\boldsymbol{V}^{n+1} \right)\,,
\end{align}
where $\Delta t = t^{n+1} - t^{n}$, $\boldsymbol{M}, \boldsymbol{G} \in \mathbb{R}^{3N\times3N}$ are the mass and damping coefficient matrix, respectively; the matrices $\boldsymbol{F}^{n}, \boldsymbol{S}^{n+1}  \in \mathbb{R}^{3N\times3N}$, on the other hand, denote the total external and internal forces. In particular, each entry $s^{n+1}_{i}$ of $\boldsymbol{S}^{n+1}$ is computed as
\begin{equation*}
    s^{n+1}_{i} = \sum_{j\in \mathcal{N}(i)} \kappa_{i,j} \left( \left(\boldsymbol{x}^{n+1}_{j} - \boldsymbol{x}^{n+1}_{i}\right)^\mathsf{T} \boldsymbol{\hat{d}}^{n+1}_{i,j} - l_{i,j}\right)\boldsymbol{\hat{d}}^{n+1}_{i,j}\,,
\end{equation*}
where the stiffness $\kappa_{i,j}$ and rest-length $l_{i,j}$ characterize the spring connecting particles $i$ and $j$, the set $\mathcal{N}$ contains all particles connected to $i$, and the direction vector is computed as
\begin{equation*}
    \boldsymbol{\hat{d}}^{n+1}_{i,j} = \frac{\boldsymbol{x}^{n+1}_{j} - \boldsymbol{x}^{n+1}_{i}}{\|\boldsymbol{x}^{n+1}_{j} - \boldsymbol{x}^{n+1}_{i}\|}\,.
\end{equation*}
We follow the linearization proposed by Selle \etal~\cite{selle2008mass}, where the direction vector is kept fixed so that $\boldsymbol{\hat{d}^{n+1}} \to \boldsymbol{\hat{d}^{n}}$, and the internal force terms become
\begin{align}
\label{eq:spring_semi_implicit}
        s^{n+1}_{i} &= \sum_{j\in \mathcal{N}(i)} \kappa_{i,j} \left(\|\boldsymbol{x}^{n}_{j} - \boldsymbol{x}^{n}_{i}\| - l_{i,j} \right)\boldsymbol{\hat{d}}^{n}_{i,j} \\
        &+\sum_{j\in \mathcal{N}(i)} \kappa_{i,j} \Delta t \boldsymbol{D}^{n}_{i,j} \left( \boldsymbol{v}^{n+1}_{j} - \boldsymbol{v}^{n+1}_{i}\right) \nonumber\,,
\end{align}
where $\boldsymbol{D}^{n}_{i,j} = \left(\boldsymbol{\hat{d}}^{n}\right)^\mathsf{T} \boldsymbol{\hat{d}}^{n}$ is the direction matrix for particles $i$ and $j$. The first term in this expansion corresponds to the explicitly integrated elastic force, while the second term describes the damping of the spring, fixed at $\kappa_{i,j}\Delta t$, to ensure stability in this semi-implicit discretization.


\subsubsection{Integration}
Given the expansion in Eq.~\eqref{eq:spring_semi_implicit}, we can write the internal force vector as
\begin{equation}
    \boldsymbol{S}^{n+1} = \hat{\boldsymbol{S}}^{n} + \Delta t \boldsymbol{C}^{n} \boldsymbol{V}^{n+1}\,,
\end{equation}
where $\hat{\boldsymbol{S}}^{n}$ is the elastic term at $t=n$, and the connectivity matrix $\boldsymbol{C}^{n}\in\mathbb{R}^{3N\times3N}$ is composed of block matrices $\boldsymbol{c}^{n}\in\mathbb{R}^{3\times3}$, computed as
\begin{equation}
\boldsymbol{c}^{n}_{ij} = 
\begin{cases} 
      \kappa_{i,j} \boldsymbol{D}^{n}_{i,j}\,, & j\in\mathcal{N}(i)\,, \\
      -\sum_{j\in\mathcal{N}(i)} \kappa_{i,j} \boldsymbol{D}^{n}_{i,j}\,, & i=j\,, \\
      \boldsymbol{0}\in\mathbb{R}^{3\times3}\,, & \text{else}\,. 
   \end{cases}
\end{equation}
Note that, since our rigid-body ghost configuration is one-way coupled with the real particles, we can include the ghost interaction directly in the elastic term $\hat{\boldsymbol{S}}^{n}$, as well as by adding the velocity term separated as the external interaction $\kappa_{\tau} \Delta t \boldsymbol{D}^{n}_{g,i}\boldsymbol{w}^{n+1}_{i}$, and the damping $-\kappa_{\tau} \boldsymbol{D}^{n}_{g,i}\boldsymbol{v}^{n+1}_{i}$, where $\kappa_{\tau}$ is a dummy index for the \emph{integrity} $\kappa_{I}$ and \emph{angular} $\kappa_{\alpha}$ spring constants, and the ghost direction matrix $\boldsymbol{D}^{n}_{g,i}$ is defined as
\begin{equation}
    \boldsymbol{D}^{n}_{g,i} = 
    \begin{cases}
        \left(\hat{\boldsymbol{d}}_{g,i}\right)^\mathsf{T}\hat{\boldsymbol{d}}_{g,i} & \|\boldsymbol{x}^{n}_{i} - \boldsymbol{y}^{n}_{i}\| \neq 0\,, \\
        \boldsymbol{0}\in\mathbb{R}^{3\times3} & \text{else}\,.
    \end{cases}
\end{equation}
This separation of ghost terms allows to compute first the ghost rigid-body dynamics and then update the real strand dynamics using an $N\times N$ matrix, as opposed to an $(2N-1)\times (2N-1)$ matrix in a two-way coupled system. Combining all the interaction terms, we can express the implicit Euler step $\mathcal{E}$ as
\begin{align}
    \left(\boldsymbol{I}+\Delta t \boldsymbol{M}^{-1} \boldsymbol{G} - \Delta t^{2} \boldsymbol{M}^{-1}\boldsymbol{C}^{n}\right) \boldsymbol{V}^{n+1} &=\\ \boldsymbol{V}^{n} + \Delta t \boldsymbol{M}^{-1}\left(\boldsymbol{F}^{n} + \hat{\boldsymbol{S}}^{n}\right)\nonumber\,.
\end{align}
\subsection{Hair Interactions}
\label{sec:hair-hair}
We implement a two-stage hybrid Eulerian/Lagrangian approach, resembling that of \cite{mcadams2009detail}. In the first stage, we rasterize hair segments into a dynamic background Eulerian volume that moves rigidly with the mesh (as opposed to the static volume formulation of McAdams \etal \cite{mcadams2009detail}), and solve the equivalent fluid system using a FLIP/PIC scheme. Then, we transfer the resulting velocity back to the particles, effectively preconditioning the velocity vector, and resolve detailed Lagrangian collisions in a second stage. 


\subsection{Hair-Solid Collisions}
\label{sec:hair-solid}
The last step of an iteration corresponds to final velocity and position corrections to account for solid collision response. Having a pre-computed SDF $\sigma_{\text{head}} : \mathcal{R}^{3}\to\mathcal{R}$ of the head, with an associated velocity field $\boldsymbol{v}_{\text{head}} : \mathcal{R}^{3}\to\mathcal{R}^{3}$, we first check for particles that will be updated into an invalid position $\sigma_{\text{head}}(\boldsymbol{x}+\Delta t \boldsymbol{v})<0$. For these cases, we first update the velocity by
\begin{equation*}
    \boldsymbol{v}^{'} = \boldsymbol{v}_{\text{head}} + \mathsf{max}\left(0, 1-\mu \frac{\|\boldsymbol{v}_{N} - \boldsymbol{v}_{\text{head}_{N}}\|}{\|\boldsymbol{v}_{T} - \boldsymbol{v}_{\text{head}_{T}}\|} \right)\left(\boldsymbol{v}_{T}-\boldsymbol{v}_{\text{head}_{T}}\right)\,,
\end{equation*}
where $\boldsymbol{v}_{\text{head}}=\boldsymbol{v}_{\text{head}}(\boldsymbol{x}+\Delta t \boldsymbol{v})$, and the sub-indices $N$ and $T$ denote the normal and tangential components of velocities when projected on the level-set normal $\nabla \sigma_{\text{head}}(\boldsymbol{x}+\Delta t \boldsymbol{v})$. When the targeted positions of particles are still inside of the head, we pushed them further in a second stage by setting
\begin{equation}
    \boldsymbol{x}^{'} = \boldsymbol{x}+\left(\frac{\nabla \sigma_{\text{head}}}{\|\nabla \sigma_{\text{head}}\|} \sigma_{\text{head}}\right) \left(\boldsymbol{x} + \Delta t \boldsymbol{v}^{'}\right)\,.
\end{equation}

\section{Experiments}
\label{sec:results}
Our numerical procedure, including matrix decomposition and time integration scheme, is detailed in supplementary material. We present a variety of results simulated with our C++/CUDA framework, implemented as described in the previous section.
\begin{table}[ht]
\centering
\caption{Overview of the relevant parameters used in the scenes presented in this paper. Time $T$ is listed in ms. Unless otherwise mentioned, identical parameter values of $[\kappa_{L}] = 10^{6} \si{\,N\,m^{-1}}$, $[\kappa_{I}] = 10^{2} \si{\,N\,m^{-1}}$, $[\kappa_{\alpha}] = 10^{2} \si{\,N\,\text{rad}^{-1}}$ are used in all simulations. We used a grid resolution of $128^{3}$ for Eulerian computations. Peak GB for our largest asset was 0.58.}
\label{tab:sim_data}
\scalebox{0.94}{\begin{tabular}{clcccccc}
\toprule
\textbf{Fig.} &  \textbf{Scene} & \textbf{\# Strands} & \textbf{\# Particles} & $\boldsymbol{T}$\\
\midrule
\ref{fig:strand_comparison} &  Single Strand & $1$ & $30$ & $0.01$ \\
\ref{fig:wisp_comparison} &  Wisp Comparison & $480$ & $14k$ & $0.02$ \\
\ref{fig:wig_simulation} &  Wig & $ 10,422$ & $214k$ & $1.9$\\
\ref{fig:hands_comparison} & Hand Interaction & $15,000$ & $450k$ & $15.2$ \\
\ref{fig:hollow_sphere_comparison} & Hollow Sphere & $10,298$ & $308k$ & $1.8$ \\
\ref{fig:teaser} &  Wind & $14,718$ & $294k$ & $14.9$ \\
\ref{fig:teaser} &  Asymmetric Style & $7,528$  & $150k$ & $6.4$ \\
\ref{fig:long_beard} & Beard & $1,100$ & $22k$ & $0.03$ \\
\ref{fig:roller_coaster} & Roller-Coaster & $7,500$ & $225k$ & $7.4$ \\
\ref{fig:geometry_vs_physics} &  Digital Grooming & $7,905$ & $166k$ & $6.2$ \\
\bottomrule
\end{tabular}}
\end{table}

Table \ref{tab:sim_data} provides an overview of the different experiments presented throughout this section, including the values of relevant parameters. The computation times listed in Table~\ref{tab:sim_data} are measured on an up-to-date desktop computer running our simulation framework on a NVIDIA GeForce GTX 3080 Ti GPU. Unless otherwise mentioned, the assets we used were generated using the parametric hair model of \cite{he2025perm}.
\subsection{Ablation Studies}
We first study the impact of the biphasic interaction terms in our augmented mass-spring formulation.
\subsubsection{Global Features}
\begin{figure}[ht]
    \centering
    \includegraphics[width=\linewidth, keepaspectratio]{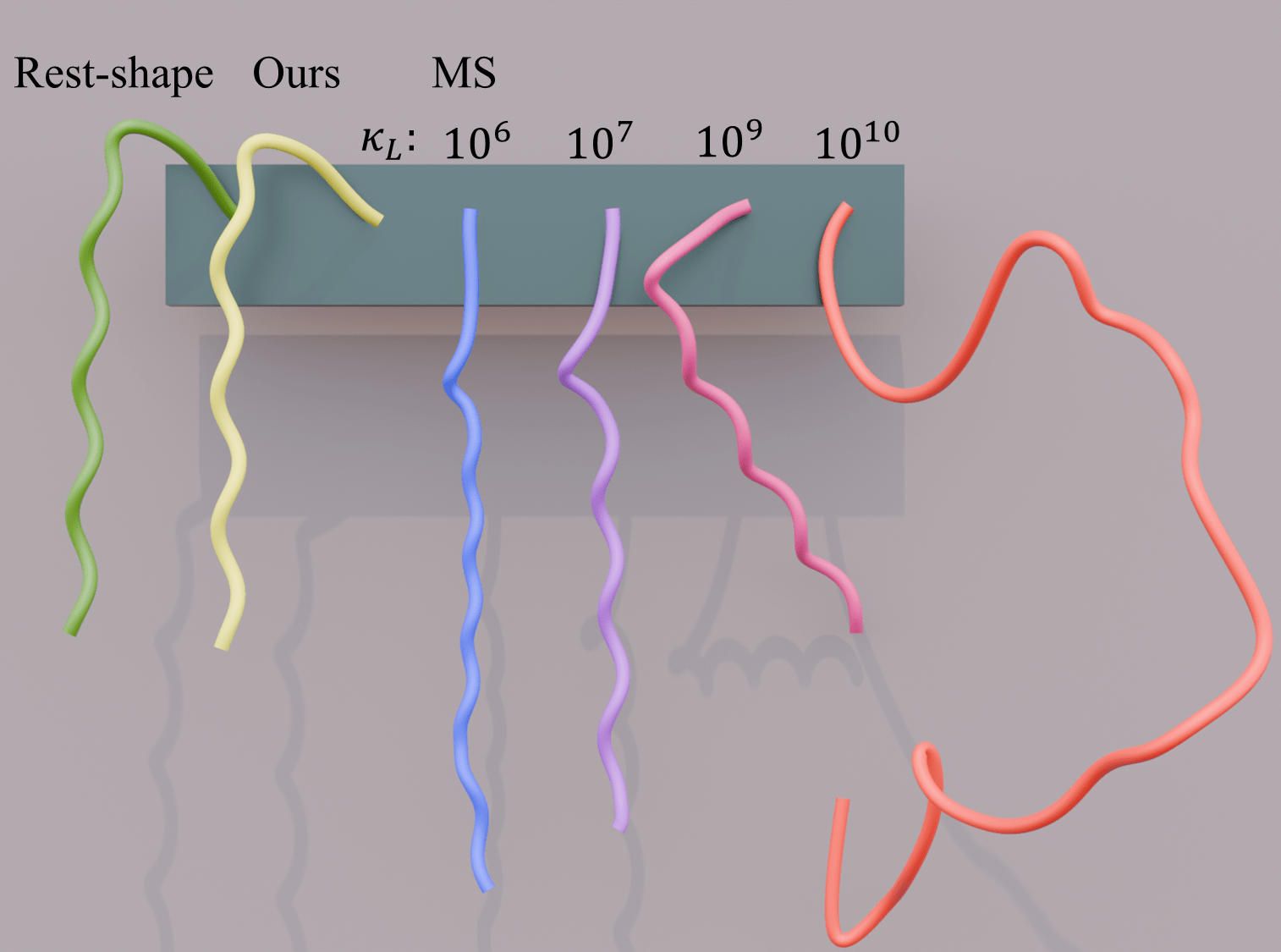}
    \caption{Simulated rest-shape of a single strand using our model and MS. Starting from a given configuration, our method preserves the global features by setting $\kappa_{L} = 10^{7}$ N${\text{m}^{-1}}$, $\kappa_{\alpha} = 0$ N${\text{rad}^{-1}}$, and $\kappa_{I} = 10^{2}$ N${\text{m}^{-1}}$, while in MS, the shape is not maintained even as $\kappa_{L}$ increases, eventually leading to instability.}
    \label{fig:strand_comparison}
\end{figure}

In Figure~\ref{fig:strand_comparison}, we present the results for a single strand simulated with $\kappa_{\alpha} = 0$ N$\text{rad}^{-1}$ in AMS, alongside simulations using MS with increasing spring tensions. Despite the local tensions being orders of magnitude higher in MS, the strand loses its shape at initialization, and, for even greater tensions, the system becomes unstable. In contrast, the \emph{integrity} tension $T_{I}$ in our formulation effectively prevents the loss of global features by directly coupling the dynamic particles to the ghost rest configuration.
\subsubsection{System Stability}
\begin{figure}[ht]
    \centering
    \includegraphics[width=\linewidth, keepaspectratio]{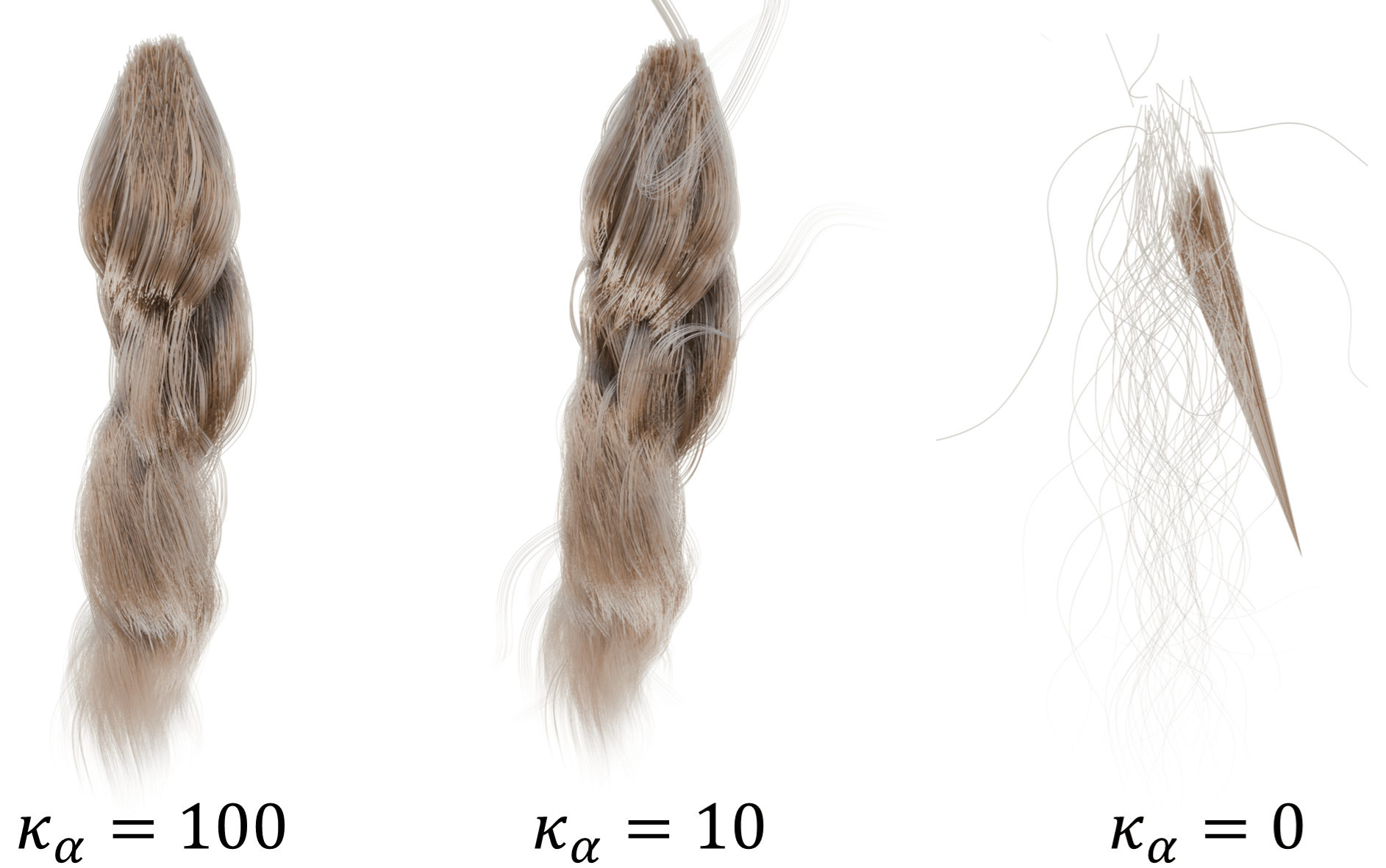}
    \caption{Wisp simulation generated using our framework, showing progressively decreasing values of the \emph{angular} coupling. As $\kappa_{\alpha}$ decreases, the simulation becomes increasingly unstable, ultimately collapsing at $\kappa_{\alpha} = 0$, where our system is equivalent to MS.}
    \label{fig:wisp_comparison}
    \vspace{-0.5cm}
\end{figure}

In Figure~\ref{fig:wisp_comparison}, we assess the impact of the \emph{angular} tension $T_{\alpha}$ in AMS by simulating a wisp of hair with our framework using $\kappa_{L} = 10^{6}$ N$\text{m}^{-1}$, $\kappa_{I} = 0$ N$\text{m}^{-1}$, and decreasing values of $\kappa_{\alpha}$. The additional degree of freedom introduced by the \emph{angular} term in our formulation enhances the stability of the system, allowing for the use of lower spring tensions and, consequently, higher time step sizes. This improvement enables real-time execution without compromising stability.
\subsubsection{Simulation Fidelity}
\begin{figure}[ht]
    \centering
    \includegraphics[width=\linewidth, keepaspectratio]{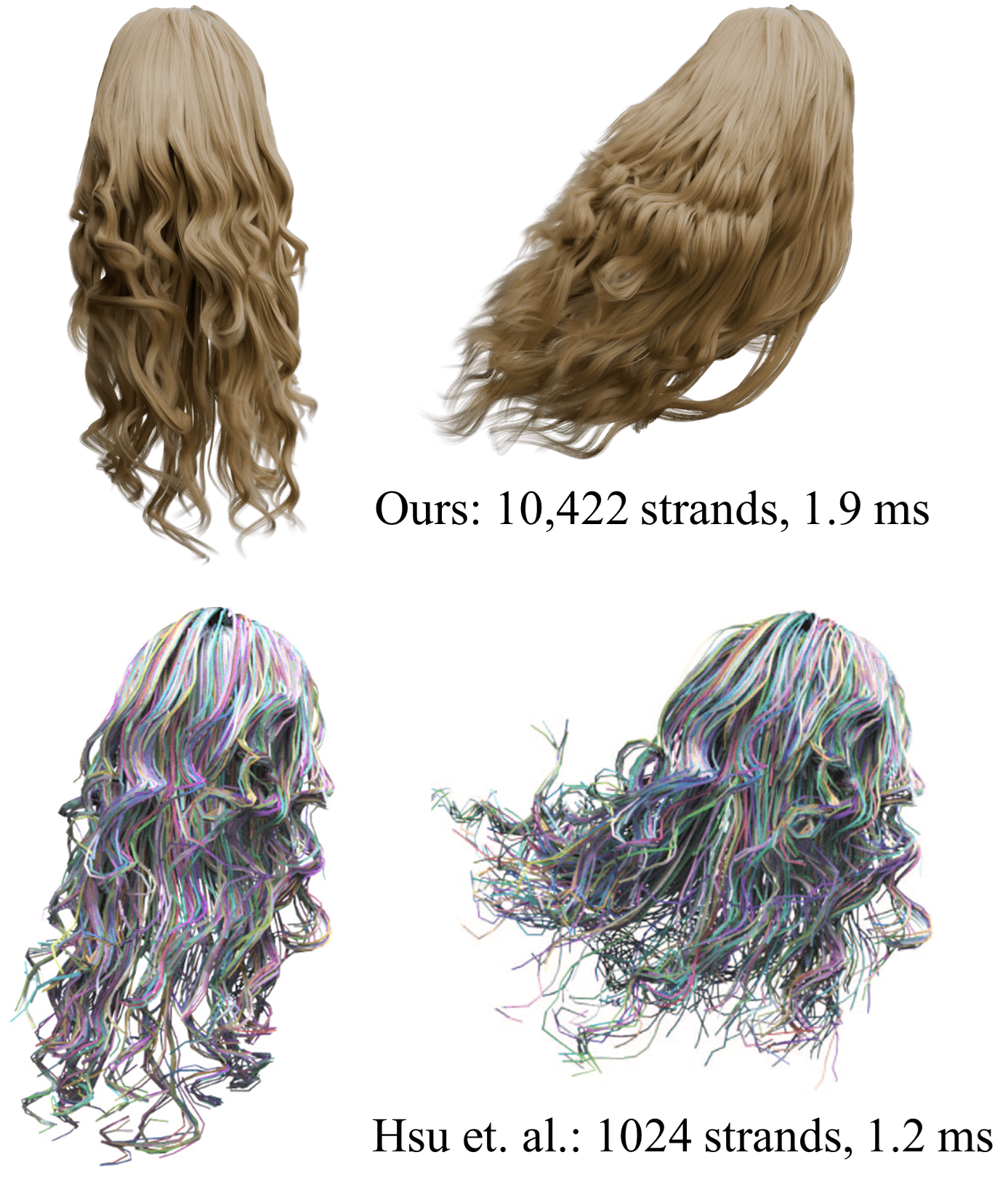}
    \caption{Qualitative comparison of the \emph{Curly-wig} experiment. Top is our simulation results. Bottom is the figure from Hsu.\etal~\cite{hsu2023sag}. }
    \label{fig:wig_simulation}
\end{figure}

To compare the simulation fidelity with the advanced hybrid Cosserat-MPM model~\cite{hsu2023sag}, we use the same real-world captured Wig model~\cite{luo2013structure} and set up a similar wind condition, running on comparable hardware, since the original paper has not released its code. As shown in Figure~\ref{fig:wig_simulation}, \cite{hsu2023sag} simulates $1024$ hair strands with a reported simulation time of $1.2$ ms per frame, while our approach requires only $0.04$ ms for the same number of strands. For higher fidelity, we increase the number of simulated strands to $10,422$ (full asset), with a per-frame time of just $1.9$ ms. Additionally, while \cite{hsu2023sag} requires specific optimizations during initialization to alleviate sagging artifacts, our approach eliminates the need for such initialization while effectively preventing sagging.
\subsection{Complex Dynamics}
We also evaluate the performance of our framework in more complex dynamic settings.
\begin{figure}[ht]
    \centering
    \includegraphics[width=\linewidth, keepaspectratio]{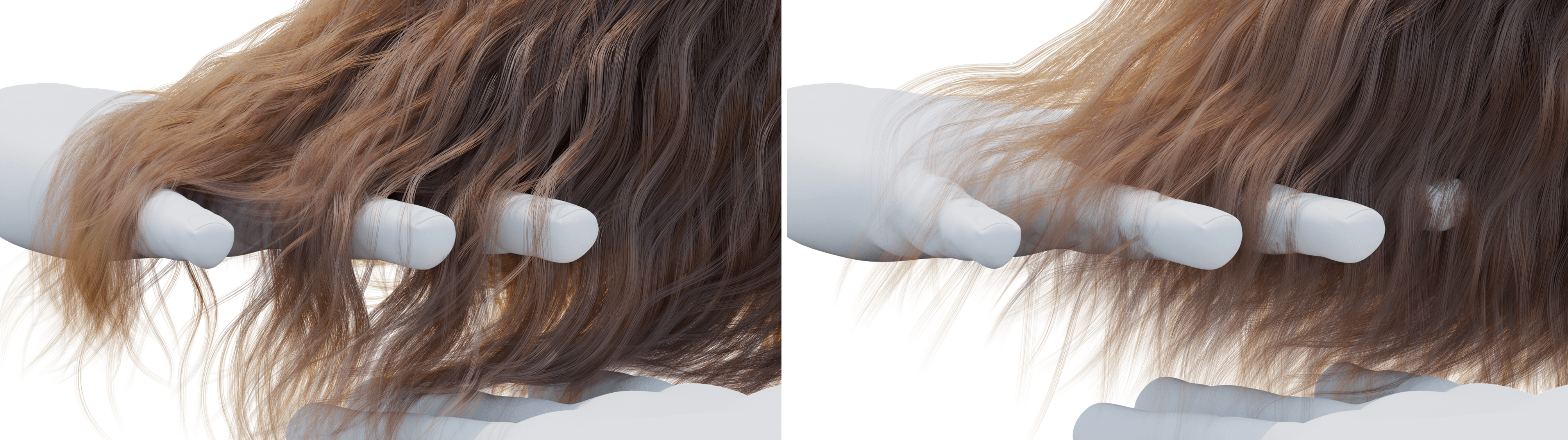}
    \caption{Hair-hand interaction simulated with our framework using 15k strands (left) and 128 guides with LHS interpolation (right). Unlike interpolation-based methods, our approach handles dense, contact-rich scenarios in real-time without artifacts.}
    \label{fig:hands_comparison}
\end{figure}
\begin{figure*}[ht]
    \centering
    \includegraphics[width=\textwidth, keepaspectratio]{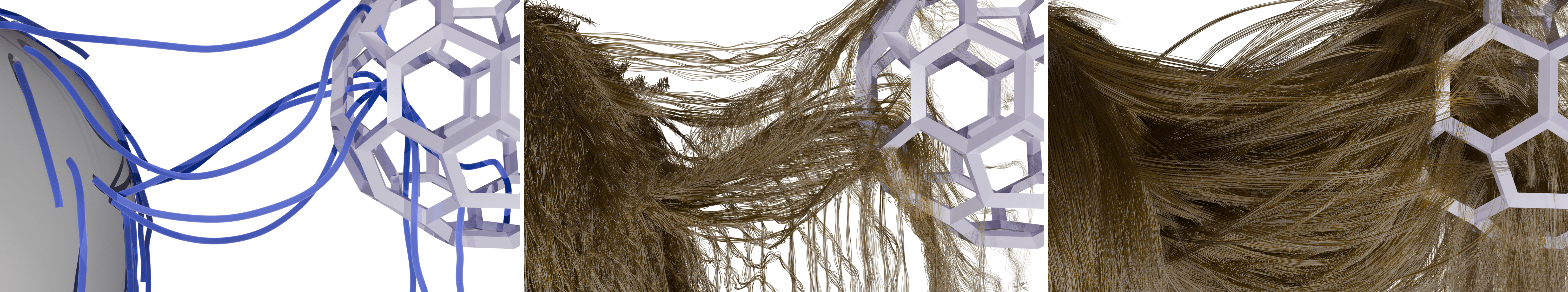}
    \caption{Simulation of hair interacting with a hollow sphere, using $128$ guiding strands (left) with interpolation based on the neural approach from \cite{shen2023ct2hair} (middle), and $10,298$ fully simulated strands with AMS (right). The neural method increases the hair count without considering the updated rest pose of the head or the introduction of new geometry in the scene (the sphere), resulting in interpolation artifacts and object penetration. In contrast, our framework accurately models the complex interaction between hair and additional geometry, increasing the fidelity throughout the simulation.}
    \label{fig:hollow_sphere_comparison}
\end{figure*}
\subsubsection{Object Interaction}
In order to manage the intensive computations required for complex hair models, in previous works, strand interpolation is often applied to a limited set of guide strands. However, this approach can lead to artifacts, particularly in scenarios involving intricate object interactions. Figure~\ref{fig:hands_comparison} illustrates this effect with hair-hand interaction, where we use $15k$ fully simulated strands alongside Linear Hair Skinning (LHS) interpolation over $2048$ guiding strands. We further investigate hair-solid interaction in Figure~\ref{fig:hollow_sphere_comparison}, using $10,298$ fully simulated strands and a neural interpolation model based on CT2Hair~\cite{shen2023ct2hair}. Unlike traditional methods, neural-driven approaches depend heavily on the training dataset and are constrained by its domain, often resulting in artifacts when handling new objects or unexpected motions. Additionally, these models require significant training effort, limiting flexibility in complex, dynamic environments.
\subsubsection{Non-Standard Settings}
\begin{figure}[ht]
    \centering
    \includegraphics[width=\columnwidth, keepaspectratio]{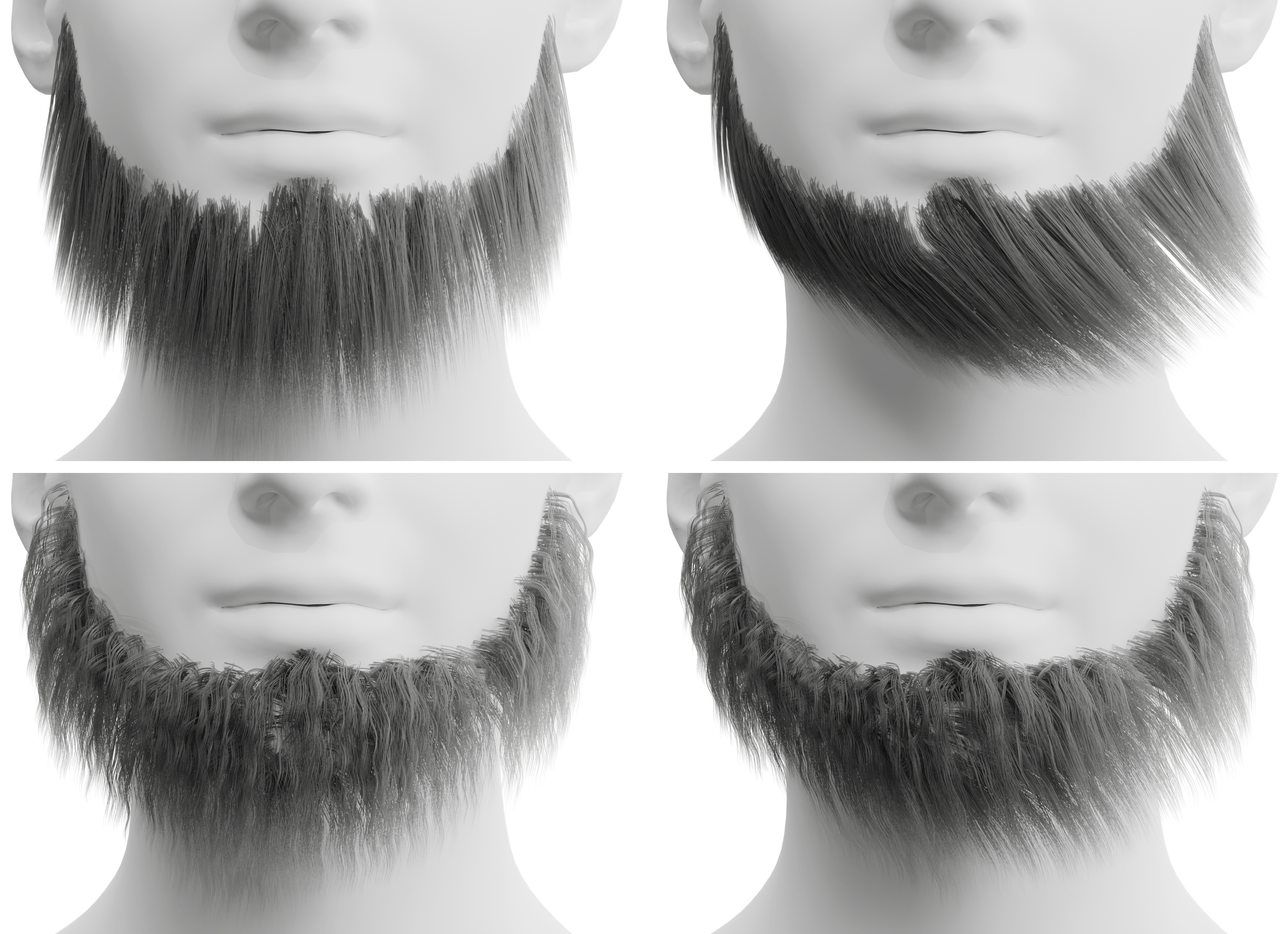}
    \caption{Time evolution (from left to right) of facial hair simulated using MS (top) and AMS (bottom). Our augmented formulation introduces key interactions that preserve the intended structure of facial hair throughout the simulation, while in MS, strands lose their overall form at initialization.}
    \label{fig:long_beard}
\end{figure}

Since MS lacks a mechanism for maintaining global structure, strands lose their intended shape across different lengths, leading to uniform behavior regardless of strand length. In contrast, the biphasic interaction in AMS enhances simulation fidelity by accounting for these variations. This advantage is especially evident in the simulation of non-standard hairstyles, as shown in Figure~\ref{fig:teaser}, middle, where we model an asymmetric style. In MS, the shorter portion of the style behaves like fur, whereas our AMS approach captures the distinct dynamics of each section, as demonstrated in the supplemental material. Another significant application is facial hair simulation. As shown in Figure~\ref{fig:long_beard}, where we simulate a beard using both MS and AMS, the initial shape of the beard is lost in MS. In contrast, our scheme preserves the overall shape and character of facial hair, while producing vivid dynamics.
\subsubsection{Extreme Forces}
\begin{figure*}[t]
    \centering
    \includegraphics[width=\textwidth, keepaspectratio]{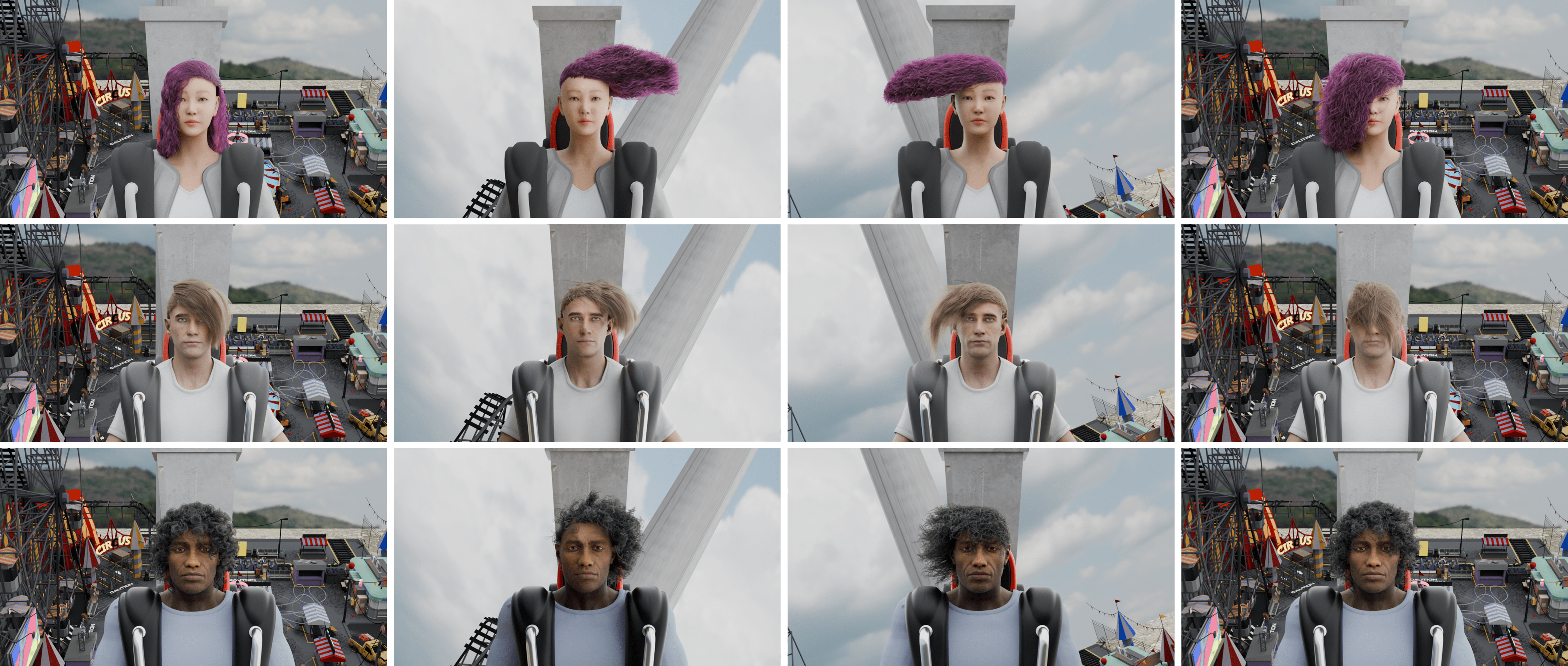}
    \caption{\fin{Time evolution (from left to right) of hair undergoing \emph{roller-coaster} dynamics simulated with our framework, shown across different styles: long (top), short (middle), and curly (bottom). The non-linear response in the biphasic interaction enables us to capture the progressive loss of global hair features when strands are subjected to intense deformation.}}
    \label{fig:roller_coaster}
\end{figure*}

Figure~\ref{fig:roller_coaster} illustrates the progressive hair shape change when strands are subjected to the intense acceleration of a roller-coaster. Our scheme is able to hold the initial hairstyle before intense motion as well as to capture the after-ride hair scramble since we explicitly parametrize non-Hookean responses on the \emph{integrity} coupling of the biphasic interaction. Moreover, our framework captures the distinct dynamical responses of varied hairstyles, such as short, long, and curly.
\subsection{Applications}
\subsubsection{Digital Grooming}
\begin{figure}[ht]
    \centering
    \includegraphics[width=\columnwidth, keepaspectratio]{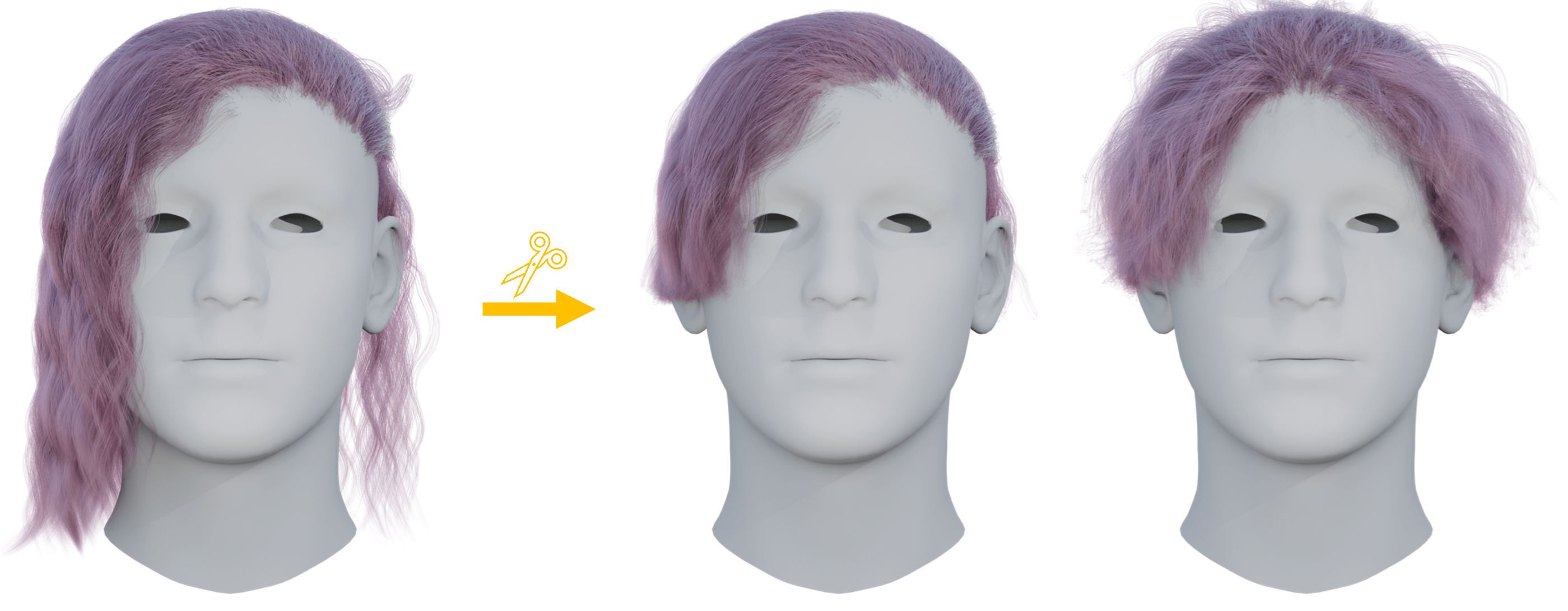}
    \caption{Starting from an initial input (left), we demonstrate purely geometrical trimming (center) versus our dynamic simulation-based trimming (right) where the hair shape changes due to the volume and length change, mimicking real-world effects.}
    \label{fig:geometry_vs_physics}
\end{figure}

Inspired by Daviet \etal~\cite{daviet2023interactive}, we explore the editing and interaction capabilities of our model by enabling diverse user inputs to transform existing hair assets into new configurations. This is illustrated in Figure~\ref{fig:geometry_vs_physics}, where we use an input (left) to compare purely geometrical trimming (center) to our dynamic simulation-based approach (right). Purely geometric approaches lack the flexibility to realistically respond to conditions like wind or complex interactions among hair strands and with external objects. 
\subsubsection{Facial Tracking Integration}
\begin{figure}[ht]
\centering
\includegraphics[width=\columnwidth, keepaspectratio]{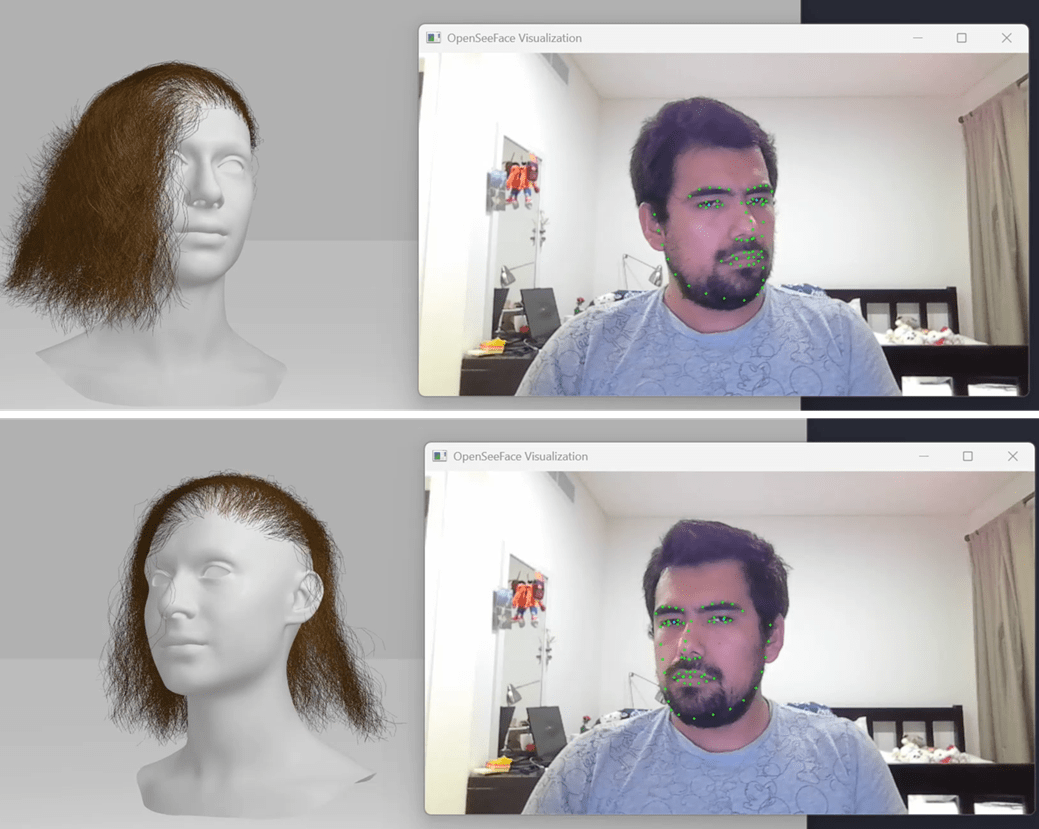}
\caption{Time evolution (from top to bottom) demonstrating the temporally coherent hair motion of a virtual avatar, driven by video-based facial tracking \cite{deng2019retinaface}. The user controls the avatar (left) via video input (right).}
\label{fig:facial_tracking_time}
\vspace{-0.6cm}
\end{figure}

We integrate facial tracking with our framework to enable digital avatars to respond dynamically to user motion, enhancing realism and interactivity. Our approach employs a video-based facial tracking system \cite{deng2019retinaface} to extract real-time head motion, which is used to drive the simulated hair. The resulting avatars exhibit temporally coherent and physically plausible behavior, as demonstrated in Figure~\ref{fig:facial_tracking_time}. Please refer to the supplemental video for a real-time demonstration.

\section{Discussion}
\label{sec:discussion}
Our AMS framework enables real-time performance of complex hair dynamics and hair-object interactions with affordable GPU memory consumption, making it feasible to incorporate high-fidelity hair simulations in games and significantly reduce the cost of hair animation in films. To demonstrate this, in the supplementary material, we showcase the real-time simulation of hair and facial hair movements for Wukong~\cite{wukong}, a hero character from a recent AAA game, running on a laptop.
\paragraph{Limitations and Future Work.}
Despite the growing popularity of neural approaches, we demonstrate in this work that data-driven methods are not a cure-all for simulation. By carefully designing physics-based models, we can achieve both high-efficiency and high-fidelity simulation effects. Future research could explore adapting AMS systems for cloth simulation. As for limitations, our method is an approximation of real-world physics, optimized for speed and visual quality, but it does not guarantee full compliance with real-world physical behavior.

{
    \small
    \bibliographystyle{ieeenat_fullname}

}

\clearpage
\appendix
\setcounter{page}{1}
\setcounter{figure}{0}
\setcounter{equation}{0}
\maketitlesupplementary

\section{Biphasic Interaction}
\label{sec:biphasic_interaction}
Despite various optimizations, DER-based methods remain computationally intensive when simulating strand dynamics, making them challenging to apply in complex interactive scenarios involving intricate hairstyles. Consequently, we focused on addressing the two primary limitations of the more efficient MS method: stability and the loss of global shape during initialization.

\begin{figure}[ht]
    \centering
    \includegraphics[width=\linewidth, keepaspectratio]{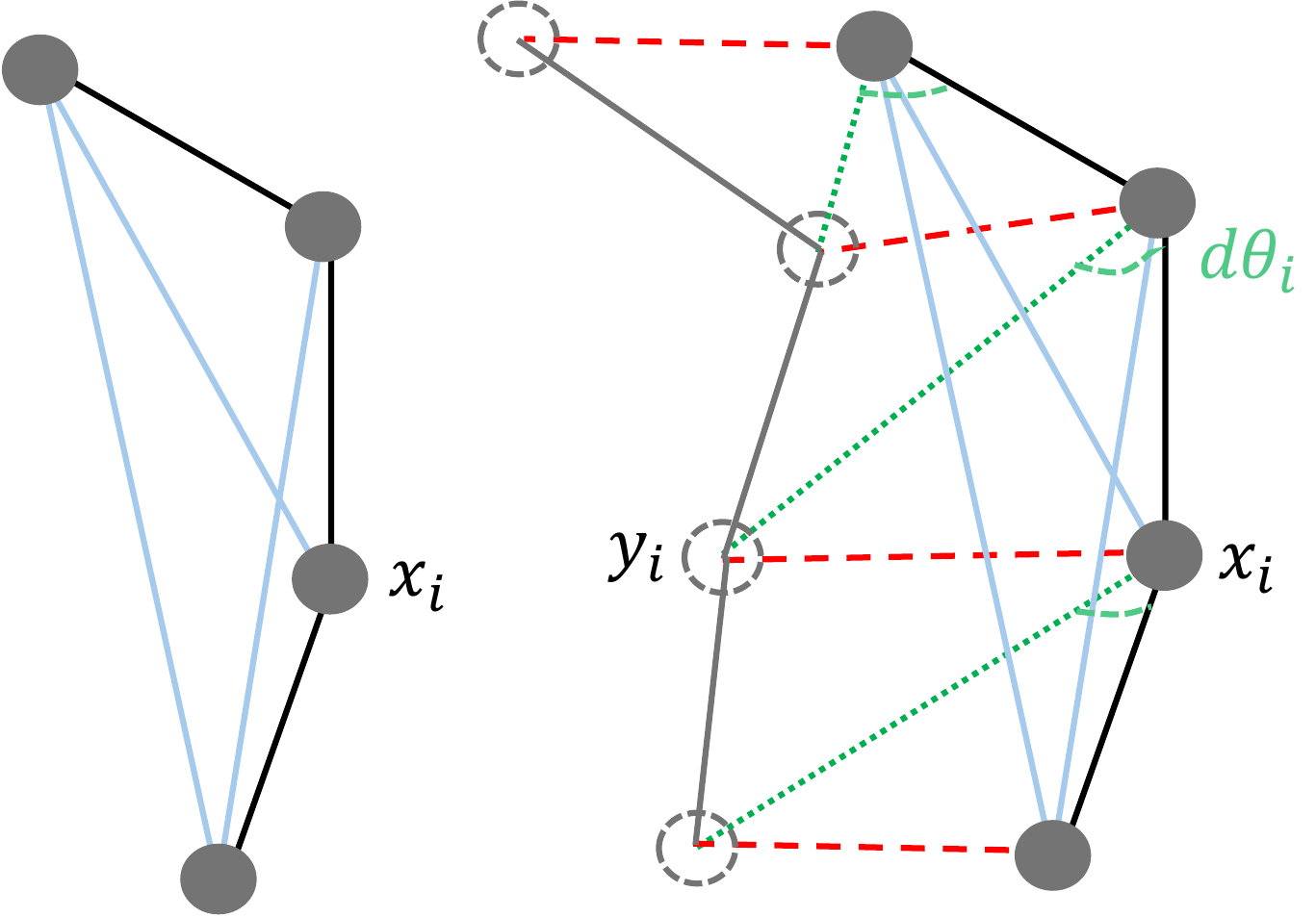}
    \caption{Schematic representation of the tetrahedra formed between consecutive real particles (left), and the additional real-ghost interaction in our formulation (right). The \emph{angular} one-way force enhances stability by preventing tetrahedron collapse when particles deviate from their original dihedral angles.}
    \label{fig:tetrahedron}
\end{figure}

Based on the detailed study of Selle \etal~\cite{selle2008mass}, stability issues in MS arise from collapsed tetrahedra formed by springs between consecutive strand particles. Because of this, our approach introduces an angular interaction with the ghost rest-shape configuration, which prevents tetrahedron collapse by maintaining an augmented stable structure based on the connections between ghost and real particles, as shown in Figure~\ref{fig:tetrahedron}.

Despite the enhanced numerical stability, the use of very stiff springs remains necessary to preserve global features, which, in practice, reintroduces instabilities unless extremely small time step sizes are employed. This constraint limits the feasibility of real-time applications. Moreover, while edge, bending, torsion, and \emph{angular} interactions maintain local shape fidelity, they fail to encode the global hair structure. To address these challenges, we encode the global features of the hair through the \emph{integrity} interaction with the rest shape, which establishes a relationship between each particle and its corresponding ghost based on the total displacement of the strand. This mechanism introduces an additional force that mitigates sagging and preserves the global shape, independently of the particle count in the discretization, by counteracting the weight of consecutive particles.

It is important to note that the two couplings we introduce for the biphasic interaction function as force perturbations to prevent tetrahedral collapse and encode global features. However, there is a potential risk that these additional forces may interfere with the fidelity of the dynamics. To mitigate this, we typically set the biphasic coupling constants several orders of magnitude lower than those of the traditional local springs, ensuring that the necessary perturbations are introduced to enhance the MS model while preserving dynamic accuracy.
\section{Numerical Integration}
Our integration procedure, summarized in Algorithm 1, updates the particle dynamics on each iteration.
\begin{algorithm}[ht]
\SetAlgoLined
\LinesNumbered
\KwIn{Current hair strands and mesh.}
\KwOut{Updated particles/mesh.}
\textbf{Procedure:}\\
|\,\,Compute $\Delta t^{'} = \nicefrac{\Delta t}{M}$\,.\\
|\,\,Define $\boldsymbol{X}^{n+1}_{0} = \boldsymbol{X}^{n}$\,.\\
|\,\,Define $\boldsymbol{V}^{n+1}_{0} = \boldsymbol{V}^{n}$\,.\\
\,\,\,\textbf{for} $i=1; \, i\leq M$ \textbf{do}\\
\,\,\,\,\,\,\,\,|\,\,Compute the intra-particle and biphasic interaction\\ 
\,\,\,\,\,\,\,\,\,\,\,terms, as described in Section \ref{sec:linearization}\,.\\
\,\,\,\,\,\,\,\,|\,\,Solve the implicit Euler step for velocity update\\
\,\,\,\,\,\,\,\,\,\,\,given by $\boldsymbol{V}^{n+1}_{i} = \mathcal{E}\,.\left(\boldsymbol{X}^{n+1}_{i-1}, \boldsymbol{V}^{n+1}_{i-1}, \boldsymbol{F}^{n}, \Delta t^{'}\right)$\\
\,\,\,\,\,\,\,\,|\,\,Update Position $\,\boldsymbol{X}^{n+1}_{i} = \boldsymbol{X}^{n+1}_{i-1} + \Delta t ^{'} \boldsymbol{V}^{n+1}_{i}$\,.\\
\,\,\,\textbf{end}\\
|\,\,Apply inextensibility constraints to modify $\boldsymbol{V}^{n+1}_{M}$ and $\boldsymbol{X}^{n+1}_{M}$\,.\\
|\,\,Rasterize velocities into dynamic background volume. \\
|\,\,Solve equivalent system through FLIP/PIC routine. \\
|\,\,Transfer velocity back to particles and resolve detailed collisions. \\
|\,\,Resolve hair-solid collisions as described in Section~\ref{sec:hair-solid}\,.
\caption{Time integration procedure of our framework.}
\label{al:main_algorithm}
\end{algorithm}

\begin{figure}[ht]
    \centering
    \includegraphics[width=\linewidth, keepaspectratio]{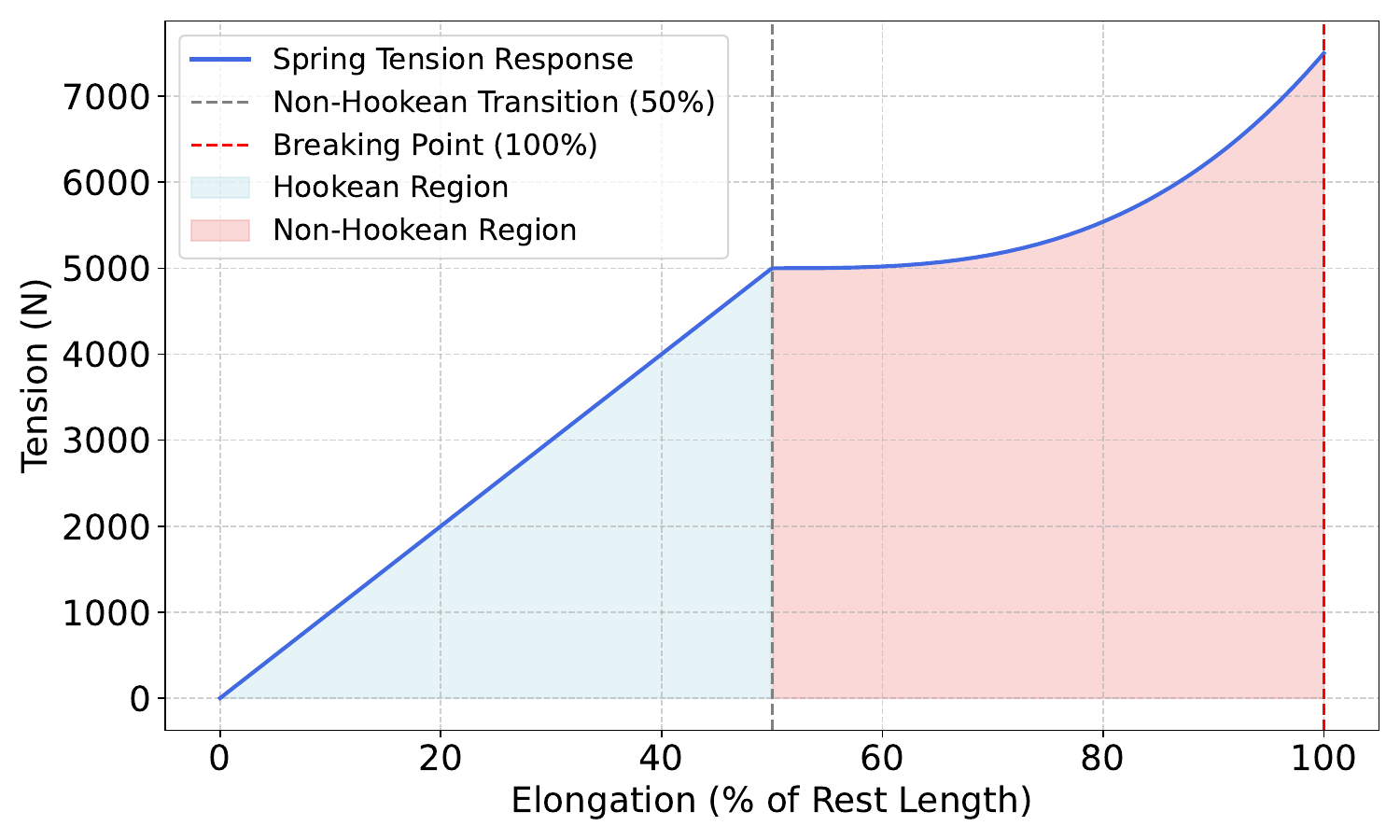}
    \caption{Parametrization plot for incorporating non-Hookean responses in the $T_{I}$ term of the biphasic coupling.}
    \label{fig:non_hookean}
    \vspace{-5mm}
\end{figure}

First, we embed the head mesh $S$ (or other solid meshes in the scene) within two 3D volumes $\Omega_{\text{Int},\text{SDF}}\in\mathbb{R}^{3}$ which we use for hair-hair interactions and SDF computation, respectively. Depending on the specific use of altitude springs and ghost configuration, the mass-spring model of Selle \etal \cite{selle2008mass} forms a banded matrix with seven to nine non-zero entries per particle, which represent the local connectivity of the system. Since we do not use two-way coupled ghosts or altitude springs, the resulting numerical system in our framework is strictly heptadiagional, which means the LU decomposition can be solved exactly using only two iterations, in a similar fashion as the solvers used in \cite{jiang2020real} and \cite{wang2019redmax}.  In general, the implicit system for a strand will have the form $\boldsymbol{A}\boldsymbol{V} = \boldsymbol{b}$, where the biphasic interaction is incorporated into $\boldsymbol{b}$, and, considering the edge, bending, and torsional degrees of freedom, the only non-zero elements in row $i$ are those at $j = i-3,\dots,i+3$. In turn , we can write the system as
\begin{align*}
    \boldsymbol{A}_{i,j}&=\begin{cases}
    -\Delta t^{2}M^{-1}_{i} \kappa_{i,j} \boldsymbol{D}_{i,j}\,, & |i-j|\leq3\,, \\
    \boldsymbol{0}\,, & \text{otherwise\,.}
    \end{cases} \\
    \boldsymbol{A}_{i,i}&=\left(1+\Delta t \boldsymbol{M}_{i}^{-1}\boldsymbol{G}_{i}\right)\boldsymbol{I} + \sum\limits_{k\in \mathcal{N}(i)} \Delta t ^{2} \boldsymbol{M}_{i}\kappa_{i,k} \boldsymbol{D}_{i,k}\,.\\
    \boldsymbol{b}_{i} &= \boldsymbol{V}^{n}_{i} + \Delta t \boldsymbol{M}^{-1}_{i}\left(\boldsymbol{F}^{n}+\hat{\boldsymbol{S}}^{n}\right)\,.
\end{align*}
This represents a linear equation in $\mathbb{R}^{3}$ and can be solved using a single forward and backward pair of sweeps. The first sweep corresponds to the decomposition $\boldsymbol{A}=\boldsymbol{L}\boldsymbol{U}$, where the strict band size of $\boldsymbol{A}$ implies that $\boldsymbol{L}_{i,j}=\boldsymbol{U}_{i,j}=\boldsymbol{0}$ for $i-j<3$ and $j-i>3$, respectively. For the other entries in the decomposition, we first do the forward sweep to compute
\begin{align*}
    \boldsymbol{L}_{i,j} &= \boldsymbol{A}_{i,j}-\sum\limits_{k=\mathsf{max}{(1,i-3)}}^{j-1} \boldsymbol{L}_{i,k}\boldsymbol{U}_{k,j}\,, \\
    \boldsymbol{V}^{'}_{i} &= \left(L_{i,i}\right)^{-1} \left[\boldsymbol{b}_{i}-\sum\limits_{j=\mathsf{max}{(0,i-3)}}^{i-1} \boldsymbol{L}_{i,j} \boldsymbol{v}^{'}_{j}\right]\,,
\end{align*}
with the intermediate vector $\boldsymbol{V}^{'} = \boldsymbol{L}^{-1}\boldsymbol{b}$. Next, the backward sweep yields
\begin{align*}
    \boldsymbol{U}_{i,j} &= \left(L_{i,i}\right)^{-1} \left[\boldsymbol{A}_{i,j}-\sum\limits_{k=\mathsf{max}{(1,j-3)}}^{i-1} \boldsymbol{L}_{i,k} \boldsymbol{U}_{k,j}\right]\,,\\
        \boldsymbol{V}_{i} &= \boldsymbol{V}^{'}_{i}-\sum\limits_{j=i+1}^{\min{(i+3,N)}} \boldsymbol{U}_{i,j}\boldsymbol{V}_{j}\,,
\end{align*}
where the final vector $\boldsymbol{V}$ is given by the relation $\boldsymbol{U}\boldsymbol{V} = \boldsymbol{V}^{'}$.
\paragraph{Non-Hookean Effects}
 In order to simulate the progressive loss of hair shape features under extreme forces, we introduce non-linear tension responses in AMS by parametrizing an elongation curve for the \emph{integrity} tension $T_{I}$ which accounts for non-Hookean behavior, as demonstrated in Figure~\ref{fig:non_hookean}.
\section{Integrity Preprocessing}
\label{sec:integrity_preprocessing}
Although our system effectively preserves global hair features, it still exhibits minor sagging effects during initialization. To mitigate this, we apply a technique similar to the \emph{gravity pre-loading} method proposed by \cite{iben2019holding}, adapted to the specific interactions relevant to our model. First we note that, at initialization, all the strand springs as well as biphasic terms are at equilibrium, so the only force acting on each particle is due to its own weight $\boldsymbol{w} = m\boldsymbol{g}$. Because of this, the initial sag stops until all the spring forces reach a new equilibrium with the total hair weight. Moreover, all of the internal Dofs and the angular interaction are given by the input configuration and then evolve dynamically. However, we can pre-process the \emph{integrity} coupling of the interaction $T_{I}$ in terms of the ghost configuration to achieve an equilibrium.

Specifically, we consider each particle $i$ with position $\boldsymbol{x}_{i}$ and its corresponding ghost at position $\boldsymbol{y}_{i}  = \boldsymbol{x}_{i} + \Delta \boldsymbol{r}_{i}$, where $\Delta \boldsymbol{r}_{i}$ is the vector joining both particles. Originally, $\Delta \boldsymbol{r}_{i} = 0$ at initialization. However, we pre-process this value to account for sagging by setting
\begin{equation}
    \boldsymbol{T}_{I} - \boldsymbol{w}_{i} = \boldsymbol{0}\,.
\end{equation}
Developing this equation we get
\begin{equation}
    \kappa_{I} d(\boldsymbol{x}_{i}, \boldsymbol{y}_{i}) \hat{\boldsymbol{r}}_{i} - \boldsymbol{w}_{i} = \kappa_{I} \|\Delta\boldsymbol{r}_{i}\| \hat{\boldsymbol{r}}_{i} - \boldsymbol{w}_{i} = \boldsymbol{0}\,.
\end{equation}
Solving this equation element-wise we finally get
\begin{equation}
    \Delta \boldsymbol{r}_{i} = \frac{m}{\kappa_{I}} \boldsymbol{g}\,.
\end{equation}
Translating the initial position of ghost particles to $\boldsymbol{y}_{i}$ = $\boldsymbol{x}_{i}+\Delta\boldsymbol{r}_{i}$ enables us to eliminate sagging at initialization.
\section{Procedural Growth}
\label{sec:procedural_growth}
\begin{figure}[ht]
    \centering
    \includegraphics[width=\columnwidth, keepaspectratio]{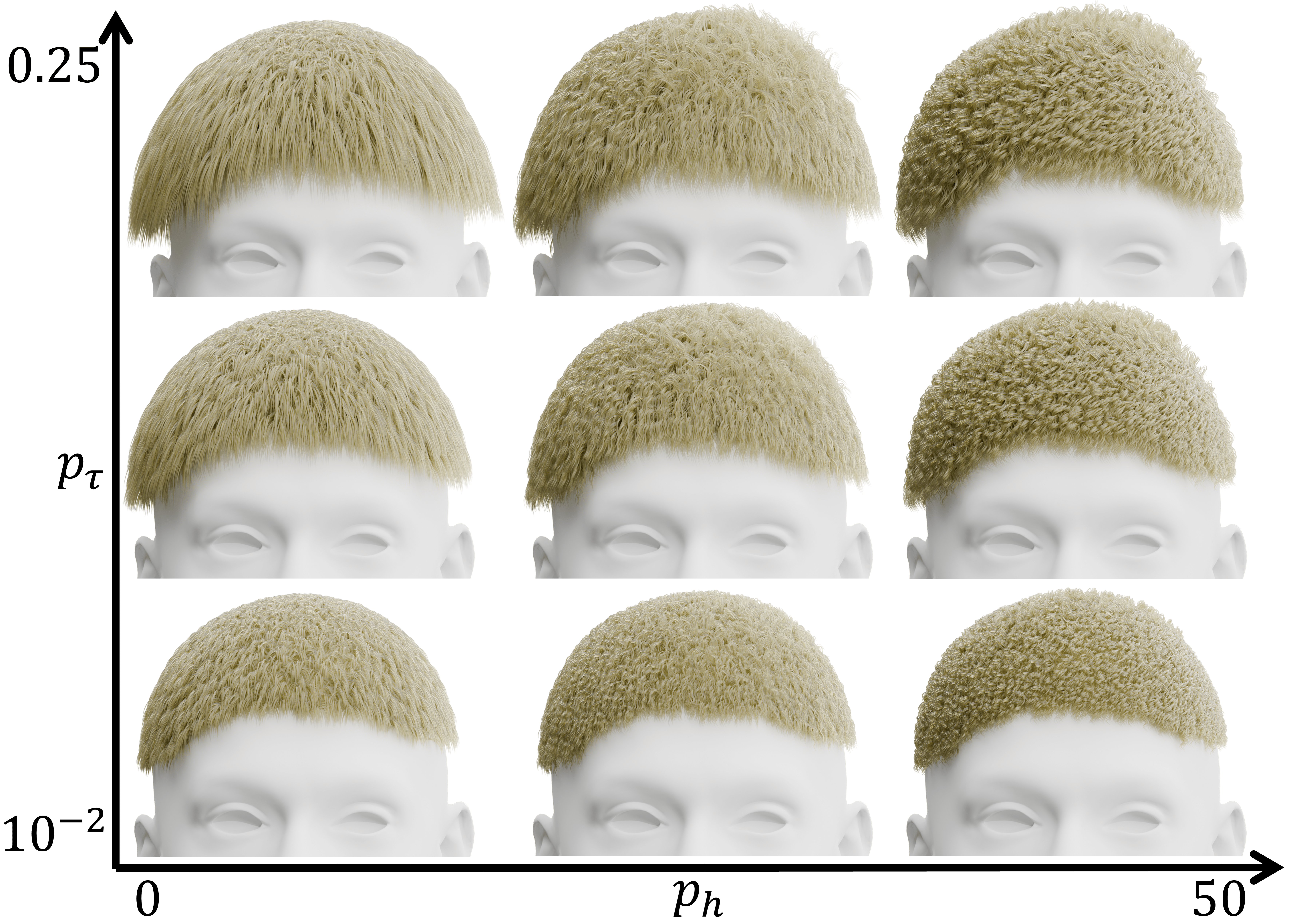}
    \caption{Parameter space exploration showing the impact of increasing values of helix radius $p_{h}$ and step size $p_{\tau}$ in our procedural hair growth module. We can control the \emph{curliness} and length of generated hair with these two parameters.}
    \label{fig:parameter_exploration}
\end{figure}
\begin{figure}[ht]
    \centering
    \includegraphics[width=\columnwidth, keepaspectratio]{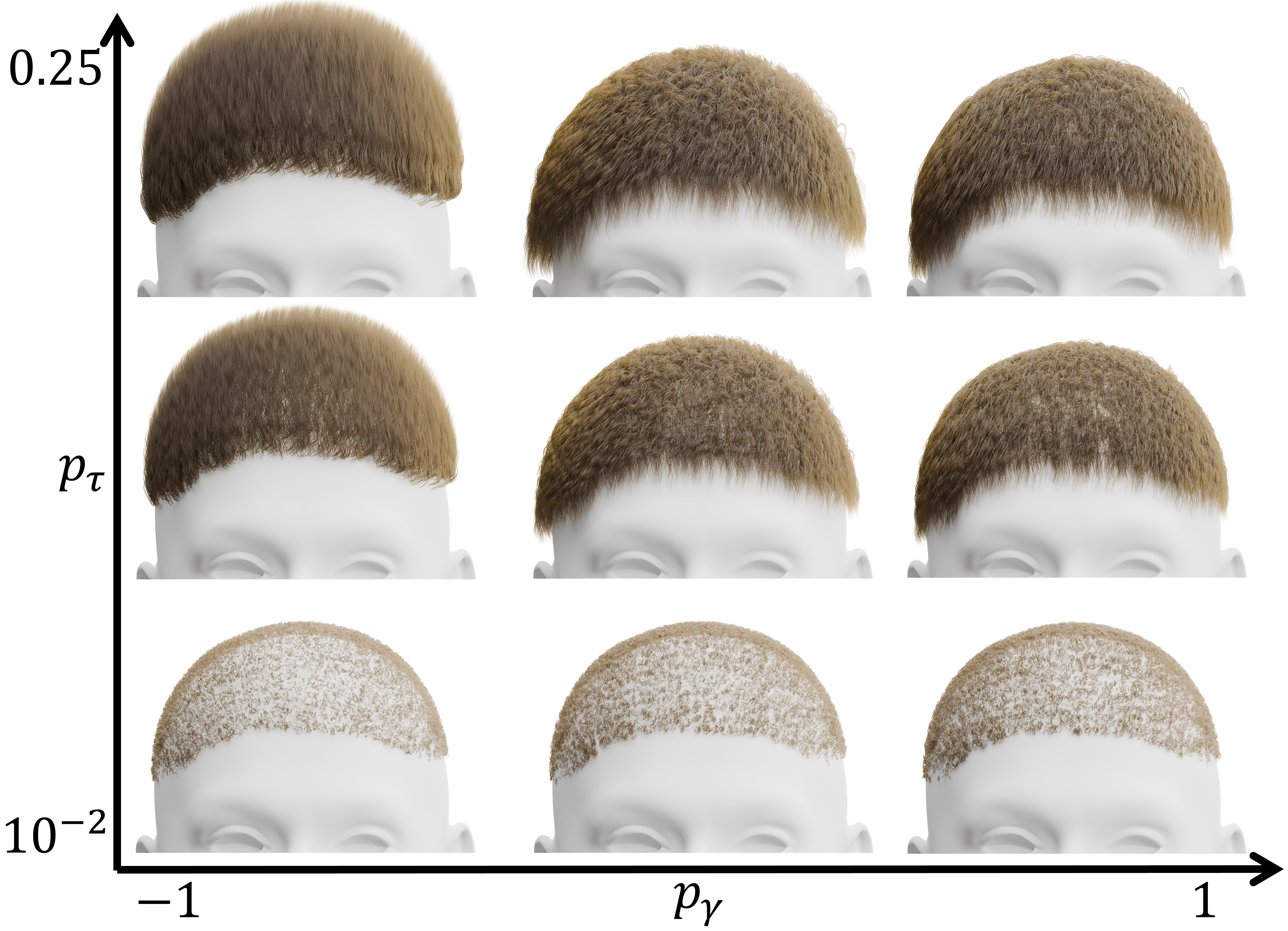}
    \caption{Parameter space exploration showing the impact of increasing values of the gravity influence parameter $p_{\gamma}$ and step size $p_{\tau}$ in our procedural hair growth module. We can control the hair deviation in the $y$ direction using different values for $p_{\gamma}$.}
    \label{fig:parameter_exploration_2}
\end{figure}

We use an heuristic approach for hair growth that is divided into two stages. First, given a pre-selected set of triangles in the mesh, we sample $p_{n}$ random root positions per triangle. Then, we compute the initial strand direction $\boldsymbol{p}^{0}_{\text{dir}}$ on each position by weighting the per-vertex normal vectors of the root using its barycentric coordinates, and adding a noise vector with entries from the distribution $\mathcal{U}(-1,1)$. During a second stage, we add sequential vertices to the strand, starting from the root. Specifically, we compute
\begin{equation}
    \boldsymbol{p}^{{i}^{'}}_{\text{dir}} = \boldsymbol{p}^{i-1}_{\text{dir}} + \boldsymbol{p}^{i-1}_{\text{grav}}\,\mathsf{max}\left(p_{\Gamma}, 1-\|\boldsymbol{p}^{i-1}_{\text{dir}}\cdot (0,1,0)\|\right)\,,
\end{equation}
where $p_{\Gamma}$ fixes the maximum particle deviation, and the procedural vector $\boldsymbol{p}^{i}_{\text{grav}}$ accounts for strand changes in the vertical direction, and is defined as
\begin{equation}
    \boldsymbol{p}^{i}_{\text{grav}} = (0,-ip_{\gamma}, 0)\,,
\end{equation}
with gravity influence parameter $p_{\gamma}$. Then, to incorporate curls into our procedural growth module, we perform an additional update step
\begin{equation}
    \boldsymbol{p}^{i}_{\text{dir}} = \boldsymbol{p}^{{i}^{'}}_{\text{dir}} + p_{\Omega} \left( \boldsymbol{p}^{{i}^{'}}_{\text{dir}}-\boldsymbol{H}^{i}\right)\,,
\end{equation}
with spiral impact factor $p_{\Omega}$, and helix vector $\boldsymbol{H}^{i}$ described by
\begin{equation}
    \boldsymbol{H}^{i} = \left((p_{h} \mathsf{cos}{\left(ip_{\text{freq}}\right)},1, p_{h} \mathsf{sin}{\left(ip_{\text{freq}} \right)}\right)\,,
\end{equation}
with helix radius $p_{h}$. We demonstrate the generation capabilities of our procedural growth scheme by performing two parameter space explorations, as shown in Figures~\ref{fig:parameter_exploration} and \ref{fig:parameter_exploration_2}. In both cases, we set $p_{\Gamma}=0.2$, $p_{\text{freq}}=1$, and $p_{\Omega}=0.017$.

\end{document}